\documentclass[twocolumn,amsmath,amssymb]{revtex4-1}
\usepackage{graphicx} 
\usepackage{dcolumn} 
\usepackage{bm} 
\usepackage{braket}
\usepackage[
colorlinks=true, 
citecolor=blue, 
urlcolor=blue, 
linkcolor=blue,
setpagesize=false, 
bookmarks=false
]{hyperref}
\usepackage[english]{babel}

\begin{document}


\title{
Pair correlations in the two-orbital Hubbard ladder: Implications on superconductivity in the bilayer nickelate La$_3$Ni$_2$O$_7$
}
\author{Tatsuya Kaneko,$^{1}$ Hirofumi Sakakibara,$^2$ Masayuki Ochi,$^{1,3}$ and Kazuhiko Kuroki$^{1}$}
\affiliation{
$^1$Department of Physics, Osaka University, Toyonaka, Osaka 560-0043, Japan \\
$^2$Advanced Mechanical and Electronic System Research Center (AMES), Faculty of Engineering, Tottori University, Tottori, Tottori 680-8552, Japan\\
$^3$Forefront Research Center, Osaka University, Toyonaka, Osaka 560-0043, Japan
}

\date{\today}


\begin{abstract}
Motivated by high-temperature superconductivity in pressurized La$_3$Ni$_2$O$_7$, we investigate the pair correlations in the two-orbital Hubbard ladder, which consists of the nearly half-filled and nearly quarter-filled orbitals. 
By employing the density matrix renormalization group method, we demonstrate that the pair correlation exhibits a power-law decay against the distance while the spin correlation decays exponentially. 
The decay exponent of the pair correlation of the nearly half-filled orbital is comparable to the exponent of the quasi-long-range superconducting correlation in the doped single-orbital Hubbard ladder, which suggests the importance of the $d_{3z^2-r^2}$ orbital in La$_3$Ni$_2$O$_7$. 
\end{abstract}

\maketitle


\section{Introduction}

Bilayer systems exhibit rich electronic properties unrealized in single-layer systems. 
A representative example is the twisted-bilayer system, where twist-angle engineering enables us to create a new platform of correlated electron physics~\cite{cao2018,kennes2021}.  
In terms of an unconventional pairing of superconductivity, bilayer and ladder lattice are intriguing structures.  
Intuitively, when interlayer couplings are strong, doped holes in a nearly half-filled bilayer prefer to form interlayer pairs in order to avoid breaking electronic rung spin-singlet bonds~\cite{dagotto1992}. 
Theoretical studies predict possible strong pairing meditated by the interlayer magnetic coupling in doped bilayer/ladder Mott insulators~\cite{dagotto1992,troyer1996,bohrdt2022}, and recently the magnetically meditated pairing has been demonstrated by ultracold atoms in the optical ladder lattice~\cite{hirthe2023}. 
Hence, bilayer and ladder systems have been considered to be promising hosts of unconventional pairing for superconductivity. 

The recent discovery of high-temperature (high-$T_{\rm c}$) superconductivity in La$_3$Ni$_2$O$_7$ under pressure~\cite{sun2023,hou2023,zhang2023_arXiv2307.14819,sakakibara2023TE,wang2023} has opened up a new avenue of investigation in the field of condensed-matter physics. 
This bilayer Ruddlesden-Popper nickelate has been theoretically proposed in 2017 as a good starting point for seeking high-$T_{\rm c}$ superconductivity by one of the present authors~\cite{nakata2017}. 
The aim in Ref.~\cite{nakata2017} was to realize in an actual material a nearly half-filled bilayer Hubbard model with the hopping between the layers $t_{\perp}$ being several times larger than the in-plane hopping $t_{\parallel}$, a model which has been suggested to exhibit high-$T_{\rm c}$ superconductivity with strong interlayer pairing~\cite{kuroki2002,maier2011,mishra2016,nakata2017,maier2019,matsumoto2020}. 
La$_3$Ni$_2$O$_7$ indeed possesses nearly half-filled $d_{3z^2-r^2}$ bands, which are largely split into bonding and antibonding bands due to $t_\perp$ within the $d_{3z^2-r^2}$ orbitals. 
However, the material deviates from an ideal bilayer Hubbard model in that the $d_{3z^2-r^2}$ orbital hybridizes with the $d_{x^2-y^2}$ orbital, which forms a nearly quarter-filled band~\cite{nakata2017,luo2023,yang2023_arXiv2309.01148}.
Another point that should be noted is that $t_\perp$ in this material is so large that the antibonding $d_{3z^2-r^2}$ band does not intersect the Fermi level, so that the nesting (in the strict sense of the term) of the Fermi surfaces arising from the bonding and antibonding $d_{3z^2-r^2}$ bands does not exist and hence cannot be the origin of superconductivity. 

In a recent paper by three of the present authors, a two-orbital bilayer (four orbitals in a unit cell) model derived from first-principles calculation was studied within the fluctuation exchange (FLEX) approximation~\cite{bickers1989}, which showed that superconductivity in  La$_3$Ni$_2$O$_7$ can still be considered as a consequence of a manifestation of the bilayer Hubbard model comprising the $d_{3z^2-r^2}$ orbitals~\cite{sakakibara2023}. 
However, FLEX is basically a weak-coupling approach and has its limitations, especially when it is applied to strongly correlated systems such as those consisting of the  $d_{3z^2-r^2}$ orbitals of La$_3$Ni$_2$O$_7$. 
In fact, there are various theoretical studies regarding the electronic structure and/or mechanism of superconductivity in La$_3$Ni$_2$O$_7$~\cite{qin2023,yang2023_PRB108.L140505,zhang2023_PRB108.165141,yang2023_PRB.108.L201108,shen2023,christiansson2023,oh2023,yang2023_PRB108.L180510,lechermann2023,liu2023,jiang2024,qu2024,wu2023,cao2023,chen2023,lu2023_arXiv2307.14965,zhang2023_arXiv2307.15276,tian2023,lu2023_arXiv2308.11195,jiang2023_arXiv2308.11614,luo2023_arXiv2308.16564,zhang2023_arXiv2309.05726,geisler2023,lange2023}, some of which suggest the importance of not only the $d_{3z^2-r^2}$ orbital but also the $d_{x^2-y^2}$ orbital~\cite{lu2023_arXiv2307.14965,oh2023,qu2024}. 

Given this background, in this article, we investigate the problem using an alternative approach that can cope with strong correlation effects with accuracy, namely, we adopt the density matrix renormalization group (DMRG) method to study a two-orbital Hubbard ladder model, a one-dimensional analog of the two-orbital bilayer Hubbard model.  
In one-dimensional ladderlike structures, a hallmark of superconductivity is given by quasi-long-range pair correlation, which decays as a power law against the distance (due to the Mermin-Wagner theorem)~\cite{jiang2019,gong2021,dolfi2015}. 
We find that when one of the two orbitals is nearly half-filling with strong $t_{\perp}$ and the other orbital is nearly quarter-filling with weak rung coupling, as expected in the bilayer nickelate, the correlation function of the rung pairing shows a power-law decay while the spin correlation exhibits an exponential decay.  
The decay exponent of the rung pairing in the nearly half-filled ladder (i.e., $d_{3z^2-r^2}$ orbital) is comparable to the exponent of the quasi-long-range superconducting order realized in the doped single-orbital Hubbard ladder~\cite{dolfi2015}.  
Our numerical demonstrations suggest that the $d_{3z^2-r^2}$ orbital that makes the strong interlayer bond in La$_3$Ni$_2$O$_7$ plays a key role in superconductivity.


\section{Model}

\begin{figure}[b]
\begin{center}
\includegraphics[width=0.8\columnwidth]{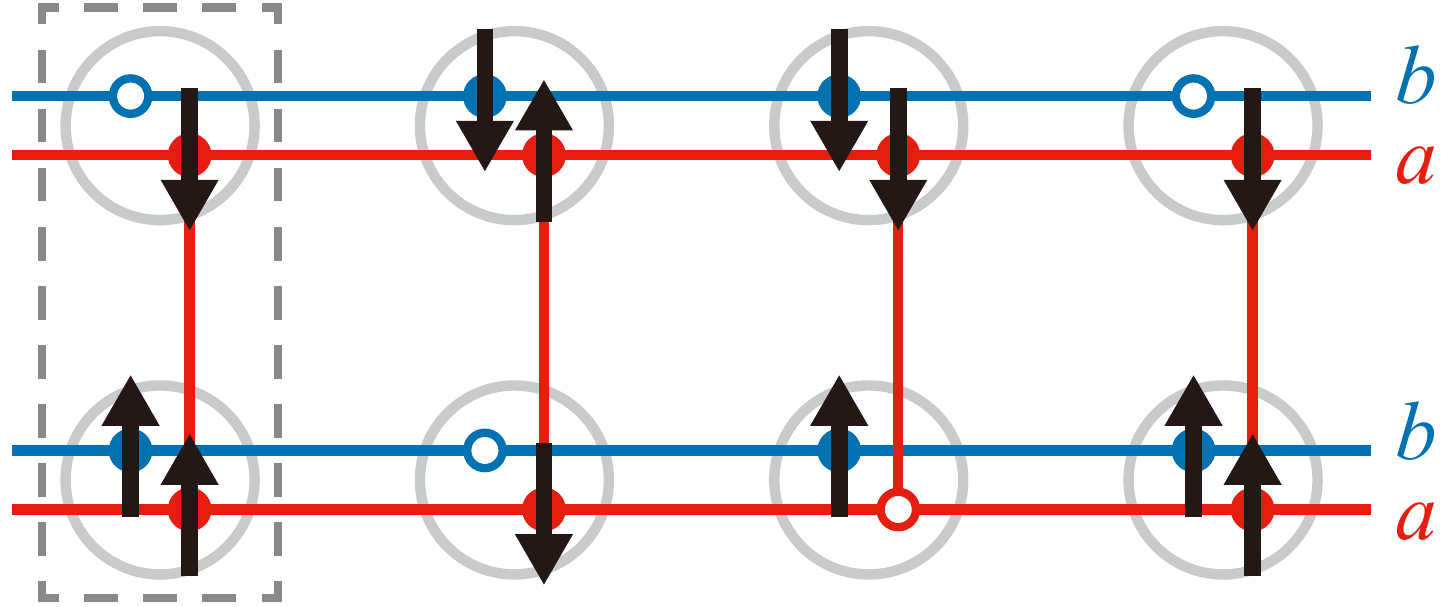}
\caption{Two-orbital Hubbard ladder at 3/8 filling. 
The dashed rectangle indicates the unit cell. 
The $a$ and $b$ orbitals correspond to the $d_{3z^2-r^2}$ and $d_{x^2-y^2}$ orbitals, respectively, in the bilayer nickelate. 
}
\label{fig1}
\end{center}
\end{figure}

To address a similar situation to the bilayer nickelate in a one-dimensional system, we consider a two-orbital Hubbard ladder at 3/8 filing, i.e., three electrons per unit cell, which consists of two sites with two orbitals on each site (see Fig.~\ref{fig1}).  
The Hamiltonian of the two-orbital Hubbard ladder reads 
\begin{align}
\hat{\mathcal{H}} =
& -\sum_{j,l} \sum_{\mu,\nu} \sum_{\sigma} t^{\mu\nu}_{\parallel} \left( \hat{c}^{\dag}_{j,l,\mu,\sigma} \hat{c}_{j+1,l,\nu,\sigma} + {\rm H.c.}\right)
\notag \\
& - \sum_{j} \sum_{\mu} \sum_{\sigma}  t^{\mu\mu}_{\perp} \left( \hat{c}^{\dag}_{j,1,\mu,\sigma} \hat{c}_{j,2,\mu,\sigma} + {\rm H.c.}\right)
\notag \\
&
+ \sum_{j,l} \sum_{\mu} \sum_{\sigma} \varepsilon_{\mu} \hat{n}_{j,l,\mu,\sigma} 
+ U \sum_{j,l}   \sum_{\mu}  \hat{n}_{j,l,\mu,\uparrow}  \hat{n}_{j,l,\mu,\downarrow}. 
\end{align}
$\hat{c}^{\dag}_{j,l,\mu,\sigma}$ ($\hat{c}_{j,l,\mu,\sigma}$) is the creation (annihilation) operator of an electron at site $j$ in chain $l$ $(=1,2)$, where $\mu$ $(=a,b)$ and $\sigma$ ($=\uparrow,\downarrow$) denote the orbital and spin degrees of freedom, respectively, and $ \hat{n}_{j,l,\mu,\sigma} = \hat{c}^{\dag}_{j,l,\mu,\sigma} \hat{c}_{j,l,\mu,\sigma}$.  
Considering this model as a one-dimensional version of that of the bilayer nickelate, orbitals $a$ and $b$ correspond to the $d_{3z^2-r^2}$ and $d_{x^2-y^2}$ orbitals, respectively.
$t^{\mu\nu}_{\parallel}$ is the hopping integral between the nearest neighboring $\mu$ and $\nu$ orbitals along the chain direction, and $t^{\mu\mu}_{\perp}$ is the interchain hopping integral of the orbital $\mu$. 
$\varepsilon_{\mu}$ is the energy level of the orbital $\mu$, and the energy-level difference is denoted by $\Delta E = \varepsilon_{b} - \varepsilon_{a}$. 
$U$ $(>0)$ is the on-site repulsive interaction.
We set $\Delta E > 0$ at 3/8 filling, where the lower-energy $a$ orbital can be nearly half-filling while the higher-energy $b$ orbital is nearly quarter-filling, as shown in Fig.~\ref{fig1}. 
In order to make correspondence to the bilayer nickelate, we assume that the $a$ (= $d_{3z^2-r^2}$) orbitals have strong interchain bonding via $t^{aa}_{\perp}$, while the interchain bonding of the $b$ (= $d_{x^2-y^2}$) orbitals is very weak (where we set $t^{bb}_{\perp} = 0$ for simplicity). 
Note that we set $t^{ab}_{\perp}=0$, because the hopping integral between the $d_{3z^2 -r^2}$ and $d_{x^2-y^2}$ orbitals along the $z$ (rung) direction is zero in the high-symmetry structure (without tilt) of the pressurized bilayer nickelate. 
On the other hand, we assume all $t^{\mu\nu}_{\parallel}$ are nonzero, where the correspondence with the bilayer nickelates requires the relation $|t^{bb}_{\parallel}| > |t^{ab}_{\parallel}| > |t^{aa}_{\parallel}|$ because the $d_{x^2-y^2}$ ($d_{3z^2-r^2}$) orbital is elongated in the in-plane (out-of-plane) direction. 
Note that our model neglects the interorbital repulsion $U'$ for simplicity.  
However, if we keep electron filling $3/4 < n_{a} < 1$ for the $a$ orbital at $U>U'$ (including the Hartree energy shift), 
$U'$ may not qualitatively change the single-site electron configuration and effective magnetic interactions.   
We will comment on the roles of Hund's coupling $J_{\rm H}$ later.  

This 3/8-filled two-orbital ladder model consists of a lightly doped Mott insulator in the $a$-orbital network and a nearly quarter-filled itinerant electron system in the $b$-orbital network (see Fig.~\ref{fig1}), coupled by the intrachain hopping $t^{ab}_{\parallel}$.
In the single-orbital Hubbard (or $t$-$J$) ladder, it is known that the lightly doped ladder is in the Luther-Emery liquid (spin-gapped) state with power-law decays of the pair and charge-density correlation functions~\cite{dolfi2015,lu2023_prb125114}. 
In the lightly doped regime, doped holes in the ladder prefer to form interchain pairs in order to avoid the destruction of electronic rung spin-singlets. 
In particular, the strong interchain coupling (optimal ratio $t_{\perp}/t_{\parallel} \sim 1.5 $~\cite{noack1995,noack1996,noack1997,sheikhan2020}) favorably leads to the quasi-long-range pair correlation overwhelming the charge-density correlation~\cite{dolfi2015,sheikhan2020}. 
On the other hand, the nearly quarter-filled single-orbital ladder is in a Tomonaga-Luttinger liquid like state, where the pair correlation is no longer dominant~\cite{lu2023_prb125114}. 
In this context we may expect that the lightly doped $a$ (i.e., $d_{3z^2-r^2}$) orbital ladder is promising for superconductivity rather than the $b$ (i.e., $d_{x^2-y^2}$) orbital network.
However, the dominance of the pair correlation in the present two-orbital model is highly nontrivial, because the $a$- and $b$- orbital ladders are coupled via $t^{ab}_{\parallel}$. 

In our calculation we set $t^{bb}_{\parallel} = t_{\rm h}$ as the unit of energy and use $t^{aa}_{\parallel} = 0.25 t_{\rm h}$ and $t^{ab}_{\parallel} = 0.5 t_{\rm h}$ corresponding to the ratios of the intralayer hoppings in La$_3$Ni$_2$O$_7$ estimated by the first-principles calculation~\cite{sakakibara2023}. 
On the other hand, we employ the interchain hoppings $t^{aa}_{\perp}=0.7t_h$ and $t^{bb}_{\perp} = 0$. 
Although the ratio $t^{aa}_{\perp}/t^{aa}_{\parallel}$ in our model is nearly half of its ratio estimated in the bilayer nickelate~\cite{sakakibara2023}, the absence of the overlap between bonding and antibonding bands is maintained in our one-dimensional model. 
This parameter choice is because, by doubling $t^{\mu\nu}_{\parallel}$ against $t^{\mu\mu}_{\perp}$ in the one-dimensional model, the width of each band becomes comparable to the corresponding bandwidth in the two-dimensional bilayer system~\cite{hopping}.  
Here, based on the intrachain component $t^{bb}_{\parallel}$ that gives the widest energy band, we set $U /t_{\rm h} = 8$ corresponding to the values suggested in the previous studies~\cite{sakakibara2023,christiansson2023}. 
Since we modify the ratio of the hopping parameters, we adjust $\Delta E$ in order to meet the electron filling of the $a$ orbital with the electron filling of the $d_{3z^2-r^2}$ orbital expected in the bilayer nickelate~\cite{sakakibara2023}.
In the present calculation, we adopt $\Delta E/t_{\rm h}=1$, which gives a reasonable electron filling of the $a$ orbital, as we shall see.
We employ the DMRG method~\cite{white1992,white1993,schollwock2011} to obtain the ground state of the model. 
Our DMRG calculations were performed using the ITensor library~\cite{ITensor}. 
We use the $L_x \times 2$ site ladder with open boundary conditions.  
The bond dimension of the DMRG calculations is up to $m=10000$, where the truncation error is of the order $10^{-6}$.
Unless otherwise specified, we plot the results for $m=10000$.


\section{Results}

\begin{figure}[b]
\begin{center}
\includegraphics[width=\columnwidth]{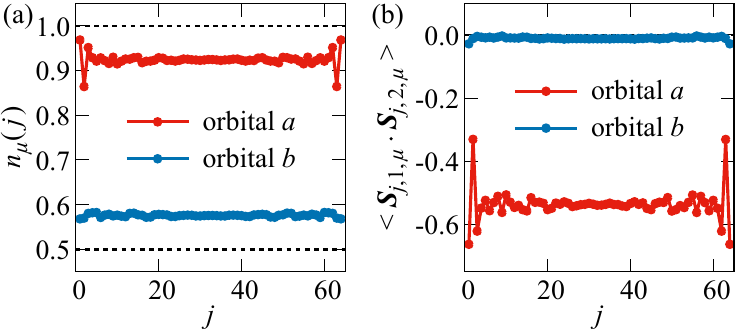}
\caption{(a) Local electron density $n_{\mu}(j)$ and (b) interchain spin correlation $\braket{\hat{\bm{S}}_{j,1,\mu} \cdot \hat{\bm{S}}_{j,2,\mu}}$ for $L_x = 64$. 
}
\label{fig2}
\end{center}
\end{figure}

In Fig.~\ref{fig2}(a) we show the calculated local electron density $n_{\mu}(j) = (1/2)\sum_{l,\sigma} \braket{\hat{n}_{j,l,\mu,\sigma}}$ for $L_x=64$. 
For the present parameter set, the $a$ orbital is close to half filling ($n_a \sim 1 - \delta$ with $\delta \sim 0.075$) while the $b$ orbital is close to quarter filling ($n_b \sim 1/2 + \delta$), in good correspondence with the bilayer nickelate. 
We do not find a clear charge-density-wave--like signature in the local electron density of the $a$ orbital $n_{a}(j)$. 
Since $t^{aa}_{\perp} > t^{aa}_{\parallel}$, two singly occupied electrons in the $a$-orbital ladder prefer to make the rung spin-singlets.  
To verify this, we plot the interchain spin correlation $\braket{\hat{\bm{S}}_{j,1,\mu} \cdot \hat{\bm{S}}_{j,2,\mu}}$ in Fig.~\ref{fig2}(b), where $\hat{S}^{\gamma}_{j,l,\mu} = (1/2) \sum_{\sigma,\sigma'} \hat{c}^{\dag}_{j,l,\mu,\sigma} (\sigma^{\gamma})_{\sigma\sigma'} \hat{c}_{j,l,\mu,\sigma'}$ and  $\sigma^{\gamma}$ is the Pauli matrix of $\gamma=x,y,z$. 
The interchain spin correlation of the $a$ orbital is negative due to the antiferromagnetic spin exchange $J^{aa}_{\perp}$ via $t^{aa}_{\perp}$. 
This implies that the $a$-orbital ladder contains the rung spin-singlet sites. 
While the presence of hole sites suppresses the spin-singlet correlation, $\braket{\hat{\bm{S}}_{j,1,a} \cdot \hat{\bm{S}}_{j,2,a}} \simeq-0.54$ in Fig.~\ref{fig2}(b) is comparable to the value $\braket{\hat{\bm{S}}_{1} \cdot \hat{\bm{S}}_{2}}=-3/4$ for a spin-singlet state of a two-site system. 
On the other hand, the interchain spin-singlet correlation of the $b$ orbitals is very weak. 
This is because we assume $t^{bb}_{\perp} =0$ as in the bilayer nickelate. 
Note that if Hund's coupling $J_{\rm H}$ is taken into account, the spin-singlet nature of the $b$ orbital may be promoted by $J_{\rm H}$ using the interorbital ferromagnetic spin alignment within the single ion and the rung spin-singlet correlation of the $a$ orbitals by $J^{aa}_{\perp}$~\cite{,lu2023_arXiv2307.14965,oh2023}. 

\begin{figure}[b]
\begin{center}
\includegraphics[width=\columnwidth]{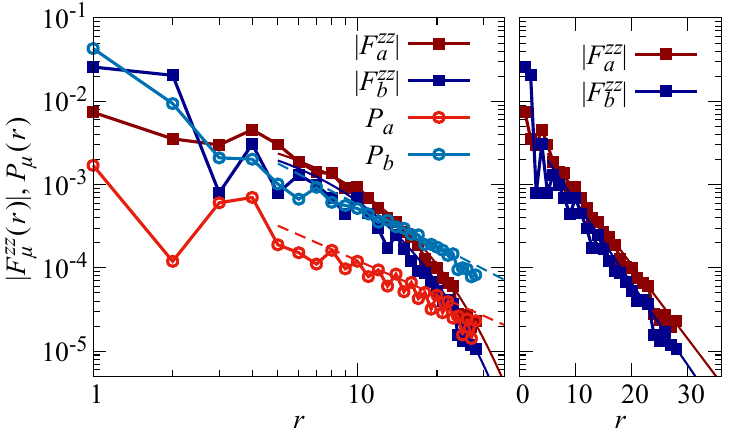}
\caption{
Left panel: Comparison of the decays of the spin-correlation function $F^{zz}_{\mu}(r)$ and pair-correlation function $P_{\mu}(r)$ for $L_x=64$ in a log-log plot.
Right panel: The same data of $F^{zz}_{\mu}(r)$ in a semilog plot. 
As a guide to the eyes, the exponential and power-law decays are indicated by the solid and dashed thin lines, respectively. 
} 
\label{fig3}
\end{center}
\end{figure}

In Fig.~\ref{fig3} we show the intrachain spin-correlation function $F^{zz}_{\mu}(r) = (1/2)\sum_{l} \braket{\hat{S}^z_{j,l,\mu} \hat{S}^z_{j+r,l,\mu}}$ in the $L_x=64$ system, where we set $j=j_{\rm ref} = 18$ as the reference site.  
Interestingly, we find that the spin correlations of both $a$ and $b$ orbitals exhibit an exponential-like decay against the distance $r$, i.e., the spin correlation cannot be dominant in the long-range part. 
This decay tendency is a signature of a spin-gapped state. 
The exponential decay of $F^{zz}_{a}(r)$ can be interpreted by the presence of the rung spin-singlet in the $a$-orbital system as shown in Fig.~\ref{fig2}(b). 
On the other hand, while the rung spin-singlet correlation of the $b$ orbital is small, the intrachain spin correlations of the $b$ orbital show an exponential decay. 
The $a$ and $b$ orbitals are entangled via the intrachain hopping $t^{ab}_{\parallel}$, which synthesizes the intrachain spin correlations and could be the origin of the exponential decay of $F^{zz}_{b}(r)$.

The tendency of the spin-singlet formation in the $a$-orbital network can be a good ingredient for a magnetically meditated interchain pairing. 
To discuss superconductivity in the two-orbital Hubbard ladder, we consider the interchain spin-singlet pair of the $\mu$ orbital given by $\hat{\Delta}_{j,\mu} = \left( \hat{c}_{j,1,\mu,\uparrow} \hat{c}_{j,2,\mu,\downarrow} - \hat{c}_{j,1,\mu,\downarrow} \hat{c}_{j,2,\mu,\uparrow}\right)/ \sqrt{2}$ and compute the pair-correlation function $P_{\mu} (r) = \braket{ \hat{\Delta}^{\dag}_{j,\mu} \hat{\Delta}_{j+r,\mu}}$. 
In Fig.~\ref{fig3} we compare the calculated $P_{\mu}(r)$ (where $j=j_{\rm ref} = 18$) with the spin-correlation function $F^{zz}_{\mu}(r)$.    
In contrast to the spin correlations, the pair correlations of both $a$ and $b$ orbitals exhibit a power-law-like decay ($\propto r^{-K_{\mu}}$). 
At short distances, $P_b(r)$ is larger than $P_a(r)$. 
This is because the $b$-orbital ladder is close to quarter filling and contains many carriers. 
In one-dimensional systems, however, the decaying behavior of the correlation at long distances is more crucial than magnitudes at short distances. 
As seen in Fig.~\ref{fig3}, the decay of $P_a(r)$ at $r > 10$ is slower than the decay of $P_b(r)$, implying that the interchain pairing of the $a$-orbital component can importantly contribute to superconductivity. 

\begin{figure}[t]
\begin{center}
\includegraphics[width=0.99\columnwidth]{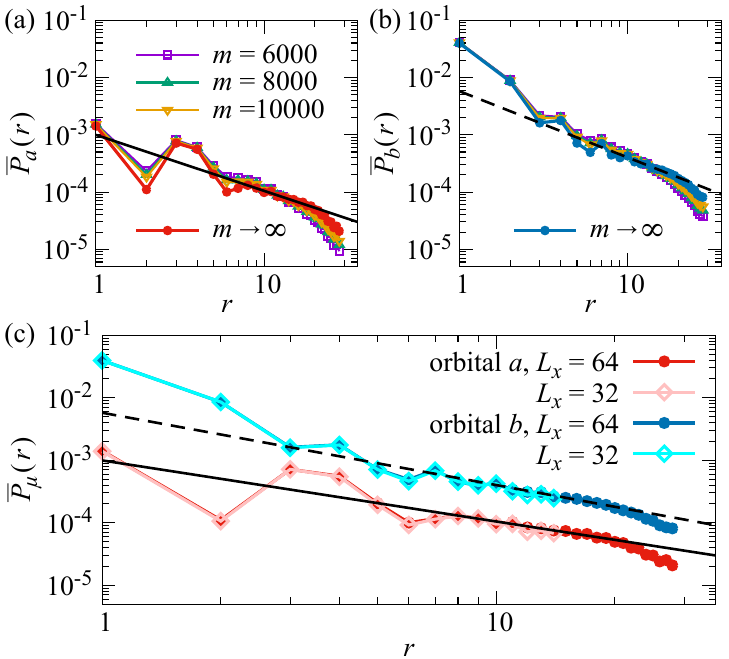}
\caption{$m$ (bond dimension) dependences of the averaged pair-correlation functions (a) $\bar{P}_{a}(r)$ and (b) $\bar{P}_{b}(r)$ for $L_x = 64$.  
The black solid and dashed lines are the fitting lines for the power-law decay of $\bar{P}_{a}(r)$ and $\bar{P}_{b}(r)$, respectively. 
(c) Comparison of the extrapolated ($m\rightarrow \infty$) pair-correlation functions, where the results for $L_x = 32$ and $L_x = 64$ are plotted. 
}
\label{fig4}
\end{center}
\end{figure}

To examine the decay of the pair correlations quantitatively, we evaluate the decay exponent $K_{\mu}$ of the correlation function $[P_{\mu}(r) \propto r^{-K_{\mu}}$]. 
Since we employ open boundary conditions in the DMRG calculations, the quantities of $P_{\mu}(r)$ have the $j_{\rm ref}$ (reference site) dependence. 
In order to get rid of the $j_{\rm ref}$ dependence as possible~\cite{noack1996,dolfi2015}, we average $P_{\mu}(r)$ over six inequivalent $j$ and $j+r$ pairs around the center of the chains, where the averaged pair correlation is defined by 
$\bar{P}_{\mu} (r) = (1/6) \sum_{s=0}^{5} \braket{ \hat{\Delta}^{\dag}_{i_0+s,\mu} \hat{\Delta}_{i_0+s+r,\mu}}$, with $i_0=(L-r+1)/2$ if $r$ is odd and  $i_0=(L-r+2)/2$ if $r$ is even. 
As discussed in Ref.~\cite{dolfi2015}, it should be noted that correlation functions calculated using a matrix-product state can reproduce a power-law decay only up to certain distances, which increase as the bond dimension $m$ is increased, while it exhibits an exponential-like decay at longer distances. 
Here, for precise evaluation we extrapolate the data of the averaged pair correlation $\bar{P}_{\mu}(r)$ with respect to the bond dimension $m$, where we use a linear fitting of $1/m$.  
Figures~\ref{fig4}(a) and \ref{fig4}(b) show the $m$ dependences of $\bar{P}_{a}(r)$ and $\bar{P}_{b}(r)$, respectively. 
The long-range part of the correlations is recovered as $m$ is increased, and the extrapolated correlations enhance the tendency of the power-law decay. 
The data drop from a monotonic power-law decay at longer distances ($r>20$) may be caused by the lack of $m$ used in our extrapolation or the finite-size effect of $L_x$. 
In Fig.~\ref{fig4}(c) we compare the extrapolated pair-correlation functions $\bar{P}_{a}(r)$ and $\bar{P}_{b}(r)$. 
We also plot the pair correlations in the $L_x = 32$ system, which show good agreement with the results in the $L_x=64$ system, indicating the convergence of the present calculation.   
The black (solid and dashed) lines in Fig.~\ref{fig4} are the fitting lines assuming $P_{\mu}(r) \propto r^{-K_{\mu}}$, where the data at $10 \le r \le 20$ in the $L_x=64$ system are used in the fitting.    
The decay exponents of the fitting lines shown in Fig.~\ref{fig4} are $K_a = 0.98$ and $K_b = 1.16$ for the $a$ and $b$ orbitals, respectively.   
The exponent $K_{a} < 1 $ in the $a$-orbital ladder is comparable to the decay exponent of the quasi-long-range pair correlation observed in the doped single-orbital Hubbard ladder~\cite{dolfi2015}. 
$K_a < K_b$ indicates that the $a$-orbital interchain pairing strongly contributes to the quasi-long-range superconducting correlation in the two-orbital Hubbard ladder. 

The present numerical results support the scenario that the $d_{3z^2-r^2}$ orbital, which is nearly half-filling, in La$_3$Ni$_2$O$_7$ importantly contributes to superconductivity. 
The interlayer pairing in the $d_{3z^2-r^2}$ orbital network may be interpreted in terms of the magnetically meditated pairing discussed in the doped single-orbital ladder and bilayer~\cite{dagotto1992,kuroki2002}.  
On the other hand, the nearly quarter-filled ladder, i.e., the $d_{x^2-y^2}$ orbital in La$_3$Ni$_2$O$_7$, also exhibits a power-law decay in the pair correlation, indicating that the pairing signature of the $d_{x^2-y^2}$ component is induced by the hybridization with the $d_{3z^2-r^2}$ orbital (via $t^{ab}_{\parallel}$). 
As discussed in the previous studies~\cite{,lu2023_arXiv2307.14965,oh2023}, Hund's coupling $J_{\rm H}$, which can promote the interlayer antiferromagnetic correlation of the $d_{x^2-y^2}$ orbitals via the antiferromagnetic rung coupling of the $d_{3z^2-r^2}$ orbitals, may enhance the contribution of the $d_{x^2-y^2}$ component to superconductivity. 
The elucidation of the effect of $J_{\rm H}$ on the pairing mechanism will be an important future extension of the present study.


\section{Summary}

We have studied the pair correlations in the two-orbital Hubbard ladder at 3/8 filling, a one-dimensional analog of the bilayer Hubbard models recently proposed for mimicking La$_3$Ni$_2$O$_7$ under pressure. 
Our DMRG calculations demonstrated that the pair correlations exhibit a power-law decay while the spin correlations decay exponentially.  
The decay exponent of the pair-correlation function is comparable to the exponent of the quasi-long-range superconducting order.   
Our numerical results suggest that the 3/8-filled two-orbital Hubbard ladder or bilayer with strong rung hopping can be a good platform for superconductivity. 
For La$_3$Ni$_2$O$_7$, our finding provides support for the scenario that the nearly half-filled $d_{3z^2-r^2}$ orbital importantly contributes to superconductivity.


\begin{acknowledgments}
This work was supported by Grants-in-Aid for Scientific Research from JSPS, KAKENHI Grants No.~JP20H01849 (T.K.), No.~JP22K03512 (H.S.), and No.~JP22K04907 (K.K.).  
The computing resource is supported by the supercomputer system (system-B) in the Institute for Solid State Physics, the University of Tokyo, and the supercomputer of Academic Center for Computing and Media Studies (ACCMS), Kyoto University.
The DMRG calculations were performed using the ITensor library~\cite{ITensor}. 
\end{acknowledgments}


\bibliography{References}

\begin{thebibliography}{58}
\bibitem [{hop()}]{hopping}%
  \BibitemOpen
  \href@noop {} {}\bibinfo {note} {If $t^{aa}_{\perp} = 0.7$ eV, corresponding
  to the value estimated in La$_3$Ni$_2$O$_7$, the values of
  $t^{\mu\nu}_{\parallel}$ (e.g., $t^{bb}_{\parallel} = 1$ eV) are roughly
  twice as large as the values of the intralayer hoppings in La$_3$Ni$_2$O$_7$
  and hence provide a comparable bandwidth with the material, since it is
  essentially a two-dimensional system.}\BibitemShut {Stop}%
\end{thebibliography}

\end{document}